\documentstyle[11pt]{article}
\begin{document}
\begin{titlepage}
\begin{center}
{\Large \bf Large Volume Axionic Swiss-Cheese Inflation}
\vskip 0.1in { Aalok Misra\footnote{e-mail: aalokfph@iitr.ernet.in} and
Pramod Shukla\footnote{email: pmathdph@iitr.ernet.in}\\
Department of Physics, Indian Institute of Technology,
Roorkee - 247 667, Uttaranchal, India}
\end{center}
\thispagestyle{empty}
\begin{abstract}
Continuing with the ideas of (section 4 of)\cite{SwissCheeseissues}, after inclusion of perturbative and non-perturbative $\alpha^\prime$ corrections to the K\"{a}hler potential and ($D1$- and $D3$-) instanton generated superpotential, we show the possibility of slow-roll axionic inflation in the large volume limit of Swiss-Cheese Calabi-Yau orientifold compactifications of type IIB string theory. We also include one- and two-loop corrections to the K\"{a}hler potential but find the same to be subdominant to the (perturbative and non-perturbative) $\alpha^\prime$ corrections. The NS-NS axions provide a flat direction for slow-roll inflation to proceed from a saddle point to the nearest dS minimum.
\end{abstract}
\end{titlepage}

\section{Introduction}

The embedding of inflation in string theory has been a field of recent interest because of several attempts to construct inflationary models in the context of string theory to reproduce CMB and WMAP observations \cite{kallosh1,wmap,KKLMMT}. These Inflationary models are also supposed to be good candidates for ``testing"  string theory \cite{kallosh1}. Initially, the idea of inflation was introduced to explain some cosmological problems like horizon problem, homogeneity problem, monopole problem etc.\cite{FirstInflation,cosmoproblem,linde}. Some ``slow roll" conditions were defined (with ``$\epsilon$" and ``$\eta$" parameters) as sufficient conditions for inflation to take place for a given potential. In string theory it was a big puzzle to construct inflationary models due to the problem of stability of compactification of internal manifold, which is required for getting a potential which could drive the inflation and it was possible to {\it rethink} about the same only after the volume modulus (as well as complex structure and axion-dilaton) could be stabilized by introducing non-perturbative effects (resulting in a meta-stable dS also) \cite{KKLT}. Subsequently, several models have been constructed with different approaches such as ``brane inflation" (for example $ D3/\overline{D3}$ branes in a warped geometry, with the brane separation as the inflaton field, D3/D7 brane inflation model \cite{KKLMMT,kesav,braneinflation}) and ``modular inflation" \cite{ book,alphacorrection,kahlerinflation}, but all these models were having the so called $\eta$- problem which was argued to be solved by fine tuning some parameters of these models. The models with multi scalar fields (inflatons) have also been proposed  to solve the $\eta$ problem \cite{Assisted}. Meanwhile in the context of type IIB string compactifications, the idea of ``racetrack inflation" was proposed by adding an extra exponential term with the same  K\"{a}hler modulus but with a different weight in the expression for the superpotential (\cite{pillado2}). This was followed by ``Inflating in a better racetrack"  proposed by Pillado et al \cite{pillado1} considering two K\"{a}hler moduli in superpotential; it was also suggested that inflation may be easier to achieve if one considers more (than one) K\"{a}hler moduli. The potential needs to have a flat direction which provides a direction for the inflaton to inflate. For the multi-K\"{a}hler moduli, the idea of treating the ``smaller" K\"{a}hler modulus as inflaton field was also proposed \cite{kahlerinflation,kahler}. Recently, ``axionic inflation" in the context of type IIB compactifications shown by Grimm and Kallosh et al \cite{Grimm,AxionInflation}, seems to be of great interest for stringy inflationary scenarios \cite{AxionInflation}. In \cite{SwissCheeseissues}, the authors had shown the possibility of getting a dS vacuum {\it without} the addition of $\overline{D3}$-branes as in KKLT scenarios \cite{KKLT}, in type IIB ``Swiss Cheese" Calabi-Yau (See \cite{SwissCheese}) orientifold compactifications in the large volume limit. In this note, developing further on this idea, we propose the possibility of axionic inflation in the same model.

The plan of the paper is as follows. In section {\bf 2}, we review the discussion of \cite{SwissCheeseissues} pertaining to obtaining a dS minimum without the addition of $\overline{D3}$-branes. We also include a discussion on one-loop and two-loop corrections to the K\"{a}hler potential. In section {\bf 3}, we discuss the possibility of getting axionic inflation with the NS-NS axions providing the flat direction for slow roll inflation to proceed starting from a saddle point and proceeding towards the nearest dS minimum. Finally, in section {\bf 4}, apart from a summary of results, we we show that it is possible to get the number of e-foldings to be 60.

\section{Getting dS Minimum Without $\overline{D3}$-Branes}

In this section, we summarize section {\bf 4} of \cite{SwissCheeseissues} pertaining to getting a de Sitter minimum without the addition of anti-$D3$ branes in type IIB ``compactifications" in the large volume limit, of orientifolds of the following two-parameter Swiss Cheese Calabi-Yau obtained as a resolution of the degree-18 hypersurface in ${\bf WCP}^4[1,1,1,6,9]$:
\begin{equation}
\label{eq:hypersurface}
x_1^{18} + x_2^{18} + x_3^{18} + x_4^3 + x_5^2 - 18\psi \prod_{i=1}^5x_i - 3\phi x_1^6x_2^6x_3^6 = 0.
\end{equation}
We also include a discussion on the inclusion of one- and two-loop corrections to the K\"{a}hler potential and show that two-loop contributions are subdominant w.r.t. one-loop corrections and the one-loop corrections are sub-dominant w.r.t. the perturbative and non-perturbative $\alpha^\prime$ corrections.

The type IIB Calabi-Yau orientifolds containing O3/O7-planes considered involve modding out by $(-)^{F_L}\Omega\sigma$ where
${\cal N}=1$ supersymmetry requires $\sigma$ to be a holomorphic and isometric involution:
$\sigma^*(J)=J,\ \sigma^*(\Omega)=-\Omega$. Writing the complexified K\"{a}hler form
$-B_2+iJ=t^A\omega=-b^a\omega_a+iv^\alpha\omega_\alpha$ where $(\omega_a,\omega_\alpha)$ form canonical
bases for ($H^2_-(CY_3,{\bf Z}), H^2_+(CY_3,{\bf Z})$), the $\pm$ subscript
indicative of being even/odd under $\sigma$, one sees that in the large volume limit of $CY_3/\sigma$,
contributions from large $t^\alpha=v^\alpha$ are exponentially suppressed, however the contributions
from $t^a=-b_a$, are not. Note that it is understood that $a$ indexes the {\bf real} subspace of {\bf real} dimensionality $h^{1,1}_-=2$; the  {\bf complexified} K\"{a}hler moduli correspond to $H^{1,1}(CY_3)$ with {\bf complex} dimensionality $h^{1,1}=2$ or equivalently real dimensionality equal to 4. So, even though $G^a=c^a-\tau b^a$ (for real $c^a$ and $b^a$ and complex $\tau$) is complex, the number of $G^a$'s is indexed by $a$ which runs over the real subspace $h^{1,1}_-(CY_3)$\footnote{To make the idea more explicit, the involution $\sigma$ under which the NS-NS two-form $B_2$ and the RR two-form $C_2$ are odd can be implemented as follows. Let $z_i, {\bar z}_i, i=1,2,3$ be the complex coordinates and the action of $\sigma$ be defined as: $z_1\leftrightarrow z_2, z_3\rightarrow z_3$; in terms of the $x_i$ figuring in the defining hypersurface in equation (1) on page 2, one could take $z_{1,2}=\frac{x_{1,2}^9}{x_5}$, etc. in the $x_5\neq0$ coordinate patch. One can construct the following bases $\omega^{(\pm)}$ of real two-forms of $H^2$ even/odd under the involution $\sigma$:
\begin{eqnarray}
\label{eq:bases}
& & \omega^{(-)}=\{\sum(dz^1\wedge d{\bar z}^{\bar 2} - dz^2\wedge d{\bar z}^{\bar 1}),
 i(dz^1\wedge d{\bar z}^{\bar 1} - dz^2\wedge d{\bar z}^{\bar 2})\}\equiv\{\omega^{(-)}_1,\omega^{(-)}_2\},\nonumber\\
& & \omega^{(+)}=\{\sum i(dz^1\wedge d{\bar z}^{\bar 2} + dz^2\wedge d{\bar z}^{\bar 1}),\sum i dz^1\wedge d{\bar z}^{\bar 1}\}\equiv\{\omega^{(+)}_1,\omega^{(+)}_2\}.
\end{eqnarray}
This implies that $h^{1,1}_+(CY_3)=h^{1,1}_-(CY_3)=2$ - the two add up to give 4 which is the {\bf real} dimensionality of $H^2(CY_3)$ for the given Swiss Cheese Calabi-Yau. As an example, let us write down $B_2\in{\bf R}$ as
\begin{eqnarray}
\label{eq:Bform}
B_2 & = & B_{1{\bar 2}}dz^1\wedge d{\bar z}^{\bar 2} + B_{2{\bar 3}}dz^2\wedge d{\bar z}^{\bar 3} + B_{3{\bar 1}}dz^3\wedge d{\bar z}^{\bar 1}  + B_{2{\bar 1}}dz^2\wedge d{\bar z}^{\bar 1} + B_{3{\bar 2}}dz^3\wedge d{\bar z}^{\bar 2} + B_{1{\bar 3}}dz^1\wedge d{\bar z}^{\bar 3}\nonumber\\
& & + B_{1{\bar 1}}dz^1\wedge d{\bar z}^{\bar 1}+ B_{2{\bar 2}}dz^2\wedge d{\bar z}^{\bar 2}+ B_{3{\bar 3}}dz^3\wedge d{\bar z}^{\bar 3}.
\end{eqnarray}
Now, using
(\ref{eq:bases}), one sees that by assuming $B_{1{\bar 2}}=B_{2{\bar 3}}=B_{3{\bar 1}}=b^1$, and $B_{1{\bar 1}}=-B_{2{\bar 2}}= i b^2, B_{3{\bar 3}}=0$, one can write $B_2=b^1\omega^{(-)}_1 + b^2\omega^{(-)}_2\equiv\sum_{a=1}^{h^{1,1}_-=2}b^a\omega^{(-)}_a$.}; the divisor-volume moduli are complexified by RR 4-form axions.
As shown in \cite{Grimm}, based on the $R^4$-correction to the $D=10$ type IIB supergravity action
\cite{Green+Gutperle}
and the modular completion of ${\cal N}=2$ quaternionic geometry by summation over all $SL(2,{\bf Z})$ images
of world sheet corrections as discussed in \cite{Llanaetal},
the non-perturbative large-volume  $\alpha^\prime$-corrections that
survive the process of orientifolding of type IIB theories (to yield ${\cal N}=1$) to the
K\"{a}hler potential is given by (in the Einstein's frame):
\begin{eqnarray}
\label{eq:nonpert8}
& & K = - ln\left(-i(\tau-{\bar\tau})\right) - 2 ln\Biggl[{\cal V} + \frac{\chi}{2}\sum_{m,n\in{\bf Z}^2/(0,0)}
\frac{({\bar\tau}-\tau)^{\frac{3}{2}}}{(2i)^{\frac{3}{2}}|m+n\tau|^3}\nonumber\\
& &  - 4\sum_{\beta\in H_2^-(CY_3,{\bf Z})} n^0_\beta\sum_{m,n\in{\bf Z}^2/(0,0)}
\frac{({\bar\tau}-\tau)^{\frac{3}{2}}}{(2i)^{\frac{3}{2}}|m+n\tau|^3}cos\left((n+m\tau)k_a\frac{(G^a-{\bar G}^a)}{\tau - {\bar\tau}}
 - mk_aG^a\right)\Biggr],
\end{eqnarray}
where $n^0_\beta$ are the genus-0 Gopakumar-Vafa invariants for the curve $\beta$ and
$k_a=\int_\beta\omega_a$,  ,
and $G^a=c^a-\tau b^a$, the real RR two-form potential $C_2=c_a\omega^a$ and the real NS-NS two-form potential
$B_2=b_a\omega^a$. As pointed out in \cite{Grimm}, in (\ref{eq:nonpert8}),
one should probably sum over the orbits of the discrete
subgroup to which the symmetry group $SL(2,{\bf Z})$ reduces. Its more natural to write out the K\"{a}hler potential
and the superpotential in terms of the ${\cal N}=1$ coordinates $\tau, G^a$ and $T_\alpha$ where
\begin{equation}
\label{eq:nonpert10}
T_\alpha = \frac{i}{2}e^{-\phi_0}\kappa_{\alpha\beta\gamma}v^\beta v^\gamma - (\tilde{\rho}_\alpha - \frac{1}{2}\kappa_{\alpha ab}c^a b^b)
-\frac{1}{2(\tau - {\bar\tau})}\kappa_{\alpha ab}G^a(G^b - {\bar G}^b),
\end{equation}
$\tilde{\rho}_\alpha$ being defined via $C_4$(the RR four-form potential)$=\tilde{\rho}_\alpha\tilde{\omega}_\alpha,
\tilde{\omega}_\alpha\in H^4_+(CY_3,{\bf Z})$. The non-perturbative instanton-corrected superpotential was shown in \cite{Grimm} to be:
\begin{equation}
\label{eq:nonpert9}
W = \int_{CY_3}G_3\wedge\Omega + \sum_{n^\alpha}\frac{\theta_{n^\alpha}(\tau,G)}{f(\eta(\tau))}e^{in^\alpha T_\alpha},
\end{equation}
where the holomorphic Jacobi theta function is given as:
\begin{equation}
\label{eq:nonpert11}
\theta_{n^\alpha}(\tau,G)=\sum_{m_a}e^{\frac{i\tau m^2}{2}}e^{in^\alpha G^am_a}.
\end{equation}
In (\ref{eq:nonpert11}), $m^2=C^{ab}m_am_b, C_{ab}=-\kappa_{\alpha^\prime ab}$, $\alpha=\alpha^\prime$
corresponding to that $T_\alpha=T_{\alpha^\prime}$ (for simplicity).

Now, for (\ref{eq:hypersurface}), as shown in \cite{Denef+Douglas+Florea}, there are two divisors which
when uplifted to an elliptically-fibered Calabi-Yau, have a unit arithmetic genus (\cite{Witten}):
$\tau_s\equiv\partial_{t_1}{\cal V}=\frac{t^2_1}{2},\ \tau_b\equiv\partial_{t_2}{\cal V}=\frac{(t_1+6t_2)^2}{2}$(the subscripts ``b" and ``s" indicative of big and small divisor volumes). In (\ref{eq:nonpert10}), $\rho_s=\tilde{\rho}_1-i\tau_s$ and
$\rho_b=\tilde{\rho}_2-i\tau_b$.

To set the notations, the metric corresponding to the K\"{a}hler potential in (\ref{eq:nonpert8}), will
be given as:
\begin{equation}
\label{eq:nonpert13}
{\cal G}_{A{\bar B}}=\left(\begin{array}{cccc}
\partial_{\rho_s}{\bar\partial}_{{\bar\rho_s}}K & \partial_{\rho_s}{\bar\partial}_{{\bar\rho_b}}K
& \partial_{\rho_s}{\bar\partial}_{{\bar G}^1}K
& \partial_{\rho_s}{\bar\partial}_{{\bar G}^2}K\\
\partial_{\rho_b}{\bar\partial}_{{\bar\rho_s}}K & \partial_{\rho_b}{\bar\partial}_{{\bar\rho_b}}K
& \partial_{\rho_b}{\bar\partial}_{{\bar G}^1}K
& \partial_{\rho_b}{\bar\partial}_{{\bar G}^2}K\\
\partial_{G^1}{\bar\partial}_{{\bar\rho}_s}K & \partial_{G^1}{\bar\partial}_{{\bar\rho}_b}K &
\partial_{G^1}{\bar\partial}_{{\bar G}^1}K & \partial_{G^1}{\bar\partial}_{{\bar G}^2}K \\
\partial_{G^2}{\bar\partial}_{{\bar\rho}_s}K & \partial_{G^2}{\bar\partial}_{{\bar\rho}_b}K &
\partial_{G^2}{\bar\partial}_{{\bar G}^2}K & \partial_{G^2}{\bar\partial}_{{\bar G}^2}K
\end{array}\right),
\end{equation}
where $A\equiv\rho^{1,2},G^{1,2}$. From the K\"{a}hler potential given in (\ref{eq:nonpert8}), one can show (See \cite{SwissCheeseissues})
that the corresponding K\"{a}hler metric of (\ref{eq:nonpert13}) is given by:
\begin{eqnarray}
\label{eq:nonpert131}
& & {\cal G}_{A{\bar B}} =\nonumber\\
& & \left(\begin{array}{cccc}
\frac{1}{4}\left(\frac{1}{6\sqrt{2}}\frac{1}{\sqrt{{\bar\rho}_s - \rho_s}{\cal Y}}
+ \frac{1}{18}\frac{({\bar\rho}_s - \rho_s)}{{\cal Y}^2}\right)
&\frac{1}{144}\left(\frac{\sqrt{({\bar\rho}_s - \rho_s)({\bar\rho}_b - \rho_b)}}{{\cal Y}^2}\right)
& \frac{-ie^{-\frac{3\phi_0}{2}}\sqrt{{\bar\rho}_s - \rho_s}{\cal Z}(\tau)}{6\sqrt{2}{\cal Y}^2}
& \frac{-ie^{-\frac{3\phi_0}{2}}\sqrt{{\bar\rho}_s - \rho_s}{\cal Z}(\tau)}{6\sqrt{2}{\cal Y}^2}  \\
\frac{1}{144}\left(\frac{\sqrt{({\bar\rho}_s - \rho_s)({\bar\rho}_b - \rho_b)}}{{\cal Y}^2}\right)
& \frac{1}{4}\left(\frac{1}{6\sqrt{2}}\frac{\sqrt{{\bar\rho}_b - \rho_b}}{{\cal Y}}
+ \frac{1}{18}\frac{\sqrt{{\bar\rho}_b - \rho_b}}{{\cal Y}^2}\right)
& \frac{-ie^{-\frac{3\phi_0}{2}}\sqrt{{\bar\rho}_b - \rho_b}{\cal Z}(\tau)}{6\sqrt{2}{\cal Y}^2}
& \frac{-ie^{-\frac{3\phi_0}{2}}\sqrt{{\bar\rho}_b - \rho_b}{\cal Z}(\tau)}{6\sqrt{2}{\cal Y}^2}  \\
\frac{ik_1e^{-\frac{3\phi_0}{2}}\sqrt{{\bar\rho}_s - \rho_s}{\cal Z}({\bar\tau})}{6\sqrt{2}{\cal Y}^2}
& \frac{ik_1e^{-\frac{3\phi_0}{2}}\sqrt{{\bar\rho}_b - \rho_b}{\cal Z}({\bar\tau})}{6\sqrt{2}{\cal Y}^2}
& k_1^2{\cal X}_1 & k_1k_2{\cal X}_1 \\
\frac{ik_2e^{-\frac{3\phi_0}{2}}\sqrt{{\bar\rho}_s - \rho_s}{\cal Z}({\bar\tau})}{6\sqrt{2}{\cal Y}^2}
& \frac{ik_2e^{-\frac{3\phi_0}{2}}\sqrt{{\bar\rho}_b - \rho_b}{\cal Z}({\bar\tau})}{6\sqrt{2}{\cal Y}^2}
& k_1k_2 {\cal X}_1 & k_2^2{\cal X}_1
\end{array}\right), \nonumber\\
& &
\end{eqnarray}
where
\begin{eqnarray}
\label{eq:nonpert14}
& & {\cal Z}(\tau)\equiv \sum_c\sum_{m,n}A_{n,m,n_{k^c}}(\tau) sin(nk.b + mk.c),\nonumber\\
&& {\cal Y}\equiv {\cal V}_E + \frac{\chi}{2}\sum_{m,n\in{\bf Z}^2/(0,0)}
\frac{(\tau - {\bar\tau})^{\frac{3}{2}}}{(2i)^{\frac{3}{2}}|m+n\tau|^3} \nonumber\\
& &
- 4\sum_{\beta\in H_2^-(CY_3,{\bf Z})}n^0_\beta\sum_{m,n\in{\bf Z}^2/(0,0)}
\frac{(\tau - {\bar\tau})^{\frac{3}{2}}}{(2i)^{\frac{3}{2}}|m+n\tau|^3}cos\left((n+m\tau)k_a\frac{(G^a-{\bar G}^a)}{\tau - {\bar\tau}}
 - mk_aG^a\right),\nonumber\\
& & {\cal X}_1\equiv\frac{\sum_c\sum_{m,n\in{\bf Z}^2/(0,0)}e^{-\frac{3\phi_0}{2}}|n+m\tau|^3|
A_{n,m,n_{k^c}}(\tau)|^2cos(nk.b + mk.c)}{{\cal Y}}
 \nonumber\\
& & + \frac{|\sum_c\sum_{m,n\in{\bf Z}^2/(0,0)}e^{-\frac{3\phi_0}{2}}|n+m\tau|^3A_{n,m,n_{k^c}}(\tau)sin(nk.b + mk.c)|^2}{{\cal Y}^2},
\nonumber\\
& & A_{n,m,n_{k^c}}(\tau)\equiv \frac{(n+m\tau)n_{k^c}}{|n+m\tau|^3}.
\end{eqnarray}

The inverse metric is given as:
\begin{equation}
\label{eq:nonpert15}
{\cal G}^{-1}=\left(\begin{array}{cccc}
({\cal G})^{\rho_s{\bar\rho_s}} & ({\cal G})^{\rho_s{\bar\rho_b}} &
({\cal G})^{\rho_s{\bar G^1}} & 0 \\
\overline{({\cal G})^{\rho_s{\bar\rho_b}}} & ({\cal G})^{\rho_b{\bar\rho_b}} & (
{\cal G})^{\rho_b{\bar G^1}} & 0 \\
\overline{({\cal G})^{\rho_s{\bar G^1}}} & \overline{({\cal G})^{\rho_b{\bar G^1}}} & \frac{1}{(k_1^2-k_2^2){\cal X}_1}
& \frac{k_2}{(k_1k_2^2-k_1^3){\cal X}_1} \\
0 & 0 & \frac{k_2}{(k_1k_2^2-k_1^3){\cal X}_1} & \frac{1}{(k_1^2-k_2^2){\cal X}_1}
\end{array}\right),
\end{equation}
where
\begin{eqnarray}
\label{eq:nonpert15ctd}
& & ({\cal G})^{\rho_s{\bar\rho_s}}=\nonumber\\
& & \frac{1}{\Delta}\Biggl[
144\,{\cal Y}^2\,{\sqrt{-{\rho_s} + {\bar\rho_s}}}\,
  \biggl( 2\,{\rho_b}\,{\cal Z}^2\,{\sqrt{-{\rho_b} + {\bar\rho_b}}} \nonumber\\
  & & -
    \left( 2\,{\cal Z}^2 + e^{3\,\phi}\,{\cal X}_1\,{\cal Y}^2 \right) \,{\bar\rho_b}\,
     {\sqrt{-{\rho_b} + {\bar\rho_b}}}
    e^{3\,\phi}\,{\cal X}_1\,{\cal Y}^2\,\left( 3\,{\sqrt{2}}\,{\cal Y} +
       {\rho_b}\,{\sqrt{-{\rho_b} + {\bar\rho_b}}} \right)  \biggr)\Biggr],\nonumber\\
& & ({\cal G})^{\rho_s{\bar\rho_b}}=\frac{1}{\Delta}\Biggl[
144\,{\cal Y}^2\,\left( -2\,{\cal Z}^2 + e^{3\,\phi}\,{\cal X}_1\,{\cal Y}^2 \right) \,
  \left( {\rho_s} - {\bar\rho_s} \right) \,
  \left( {\rho_b} - {\bar\rho_b} \right)\Biggr],\nonumber\\
& & ({\cal G})^{\rho_s{\bar G^1}}=\frac{1}{\Delta}
24\,i  \,e^{\frac{3\,\phi}{2}}\,{\cal Z}\,{\cal Y}^2\,\left( {\rho_s} - {\bar\rho_s} \right) \,
  \left( 3\,{\cal Y} + {\sqrt{2}}\,{\rho_b}\,{\sqrt{-{\rho_b} + {\bar\rho_b}}} -
    {\sqrt{2}}\,{\bar\rho_b}\,{\sqrt{-{\rho_b} + {\bar\rho_b}}}
    \right)\nonumber\\
& & ({\cal G})^{\rho_b{\bar\rho_b}}=\nonumber\\
& & \frac{1}{\Delta}
144\,{\cal Y}^2\,\biggl[ -2\,{\rho_s}\,{\cal Z}^2\,{\sqrt{-{\rho_s} + {\bar\rho_s}}} +
    \left( 2\,{\cal Z}^2 + e^{3\,\phi}\,{\cal X}_1\,{\cal Y}^2 \right) \,{\bar\rho_s}\,
     {\sqrt{-{\rho_s} + {\bar\rho_s}}} \nonumber\\
     & & +
    e^{3\,\phi}\,{\cal X}_1\,{\cal Y}^2\,\left( 3\,{\sqrt{2}}\,{\cal Y} -
       {\rho_s}\,{\sqrt{-{\rho_s} + {\bar\rho_s}}} \right)  \biggr] \,
  {\sqrt{-{\rho_b} + {\bar\rho_b}}},\nonumber\\
& & ({\cal G})^{\rho_b{\bar G^1}}=\frac{1}{\Delta}\Biggl[
-24\,i  \,e^{\frac{3\,\phi}{2}}\,{\cal Z}\,{\cal Y}^2\,\left( 3\,{\cal Y} -
    {\sqrt{2}}\,{\rho_s}\,{\sqrt{-{\rho_s} + {\bar\rho_s}}} +
    {\sqrt{2}}\,{\bar\rho_s}\,{\sqrt{-{\rho_s} + {\bar\rho_s}}}
    \right) \,\left( {\rho_b} - {\bar\rho_b} \right)\Biggr],\nonumber\\
& & ({\cal G})^{G^1{\bar G^1}}=\frac{1}{\Delta}\Biggl[
18\,e^{3\,\phi}\,{{k_1}}^2\,{\cal X}_1\,{\cal Y}^4 -
  6\,{\sqrt{2}}\,{{k2}}^2\,{\rho_s}\,{\cal X}^2\,{\cal Y}\,
   {\sqrt{-{\rho_s} + {\bar\rho_s}}} -
  3\,{\sqrt{2}}\,e^{3\,\phi}\,{{k_1}}^2\,{\rho_s}\,{\cal X}_1\,{\cal Y}^3\,
   {\sqrt{-{\rho_s} + {\bar\rho_s}}} \nonumber\\
   & & +
  6\,{\sqrt{2}}\,{{k_2}}^2\,{\rho_b}\,{\cal X}^2\,{\cal Y}\,
   {\sqrt{-{\rho_b} + {\bar\rho_b}}} +
  3\,{\sqrt{2}}\,e^{3\,\phi}\,{{k_1}}^2\,{\rho_b}\,{\cal X}_1\,{\cal Y}^3\,
   {\sqrt{-{\rho_b} + {\bar\rho_b}}} -
  8\,{{k_2}}^2\,{\rho_s}\,{\rho_b}\,{\cal Z}^2\,
   {\sqrt{-{\rho_s} + {\bar\rho_s}}}\,
   {\sqrt{-{\rho_b} + {\bar\rho_b}}}\nonumber\\
   & &  -
  \left( 3\,{\sqrt{2}}\,e^{3\,\phi}\,{{k_1}}^2\,{\cal X}_1\,{\cal Y}^3 +
     2\,{{k_2}}^2\,{\cal Z}^2\,\left( 3\,{\sqrt{2}}\,{\cal Y} -
        4\,{\rho_s}\,{\sqrt{-{\rho_s} + {\bar\rho_s}}} \right)  \right) \,
   {\bar\rho_b}\,{\sqrt{-{\rho_b} + {\bar\rho_b}}}\nonumber\\
   & &  +
  {\bar\rho_s}\,{\sqrt{-{\rho_s} + {\bar\rho_s}}}\,
   \left( 3\,{\sqrt{2}}\,e^{3\,\phi}\,{{k_1}}^2\,{\cal X}_1\,{\cal Y}^3 -
     8\,{{k_2}}^2\,{\cal X}^2\,{\bar\rho_b}\,
      {\sqrt{-{\rho_b} + {\bar\rho_b}}} +
     2\,{{k_2}}^2\,{\cal Z}^2\,\left( 3\,{\sqrt{2}}\,{\cal Y} +
        4\,{\rho_b}\,{\sqrt{-{\rho_b} + {\bar\rho_b}}} \right)  \right)\Biggr],\nonumber\\
        & &
        \end{eqnarray}
with:
\begin{eqnarray*}
& & \Delta=
-18\,e^{3\,\phi}\,{\cal X}_1\,Y^4 + 6\,{\sqrt{2}}\,{\rho_s}\,{\cal X}^2\,{\cal Y}\,
   {\sqrt{-{\rho_s} + {\bar\rho_s}}} +
  3\,{\sqrt{2}}\,e^{3\,\phi}\,{\rho_s}\,{\cal X}_1\,{\cal Y}^3\,
   {\sqrt{-{\rho_s} + {\bar\rho_s}}}\nonumber\\
   & &  -
  6\,{\sqrt{2}}\,{\rho_b}\,{\cal X}^2\,{\cal Y}\,{\sqrt{-{\rho_b} + {\bar\rho_b}}} -
  3\,{\sqrt{2}}\,e^{3\,\phi}\,{\rho_b}\,{\cal X}_1\,{\cal Y}^3\,
   {\sqrt{-{\rho_b} + {\bar\rho_b}}} +
  8\,{\rho_s}\,{\rho_b}\,X^2\,{\sqrt{-{\rho_s} + {\bar\rho_s}}}\,
   {\sqrt{-{\rho_b} + {\bar\rho_b}}} \nonumber\\
   & & +
  \left( 3\,{\sqrt{2}}\,e^{3\,\phi}\,{\cal X}_1\,{\cal Y}^3 +
     {\cal X}^2\,\left( 6\,{\sqrt{2}}\,{\cal Y} - 8\,{\rho_s}\,{\sqrt{-{\rho_s} + {\bar\rho_s}}}
        \right)  \right) \,{\bar\rho_b}\,
   {\sqrt{-{\rho_b} + {\bar\rho_b}}} \nonumber\\
   & & -
  {\bar\rho_s}\,{\sqrt{-{\rho_s} + {\bar\rho_s}}}\,
   \left( 3\,{\sqrt{2}}\,e^{3\,\phi}\,{\cal X}_1\,{\cal Y}^3 -
     8\,{\cal X}^2\,{\bar\rho_b}\,{\sqrt{-{\rho_b} + {\bar\rho_b}}} +
     {{\cal X}}^2\,\left( 6\,{\sqrt{2}}\,{\cal Y} + 8\,{\rho_b}\,
         {\sqrt{-{\rho_b} + {\bar\rho_b}}} \right)  \right).
         \end{eqnarray*}
Now, analogous to \cite{Balaetal2}, we will work in the Large Volume Scenario (LVS) limit: ${\cal V}\rightarrow\infty,
\tau_s\sim ln {\cal V},\ \tau_b\sim {\cal V}^{\frac{2}{3}}$. In this limit, the inverse metric (\ref{eq:nonpert15})-(\ref{eq:nonpert15ctd})
simplifies to (we will not be careful about the magnitudes of the numerical factors in the following):
\begin{equation}
\label{eq:nonpert16}
{\cal G}^{-1}\sim\left(\begin{array}{cccc}
-{\cal V}\sqrt{ln {\cal V}} & {\cal V}^{\frac{2}{3}}ln {\cal V}
& \frac{-i{\cal Z}ln {\cal V}}{{\cal X}_2}& 0 \\
{\cal V}^{\frac{2}{3}}ln {\cal V} & {\cal V}^{\frac{4}{3}} &
\frac{i{\cal Z}{\cal V}^{\frac{2}{3}}}{k_1{\cal X}_2}& 0\\
 \frac{i{\cal Z} ln {\cal V}}{{\cal X}_2}
 & \frac{-i{\cal Z}{\cal V}^{\frac{2}{3}}}{k_1{\cal X}_2}
& \frac{1}{(k_1^2-k_2^2){\cal X}_1}& \frac{k_2}{(k_1k_2^2-k_1^3){\cal X}_1} \\
0 & 0 & \frac{k_2}{(k_1k_2^2-k_1^3){\cal X}_1} & \frac{1}{(k_1^2-k_2^2){\cal X}_1}
\end{array}\right),
\end{equation}
where
$${\cal X}_2\equiv \sum_c\sum_{m,n\in{\bf Z}^2/(0,0)}|n+m\tau|^3|
A_{n,m,n_{k^c}}(\tau)|^2cos(nk.b + mk.c).$$ Refer to \cite{Balaetal2} for discussion on the minus sign in the $\left(G^{-1}\right)^{\rho_s{\bar\rho}_s}$.

The K\"{a}hler potential inclusive of the perturbative (using \cite{BBHL})and non-perturbative (using \cite{Grimm}) $\alpha^\prime$-corrections and one- and two-loop corrections (using \cite{loops}) can be shown to be given by:
\begin{eqnarray}
\label{eq:nonpert81}
& & K = - ln\left(-i(\tau-{\bar\tau})\right) -ln\left(-i\int_{CY_3}\Omega\wedge{\bar\Omega}\right)\nonumber\\
 & & - 2\ ln\Biggl[{\cal V} + \frac{\chi(CY_3)}{2}\sum_{m,n\in{\bf Z}^2/(0,0)}
\frac{({\bar\tau}-\tau)^{\frac{3}{2}}}{(2i)^{\frac{3}{2}}|m+n\tau|^3}\nonumber\\
& & - 4\sum_{\beta\in H_2^-(CY_3,{\bf Z})} n^0_\beta\sum_{m,n\in{\bf Z}^2/(0,0)}
\frac{({\bar\tau}-\tau)^{\frac{3}{2}}}{(2i)^{\frac{3}{2}}|m+n\tau|^3}cos\left((n+m\tau)k_a\frac{(G^a-{\bar G}^a)}{\tau - {\bar\tau}}
 - mk_aG^a\right)\Biggr]\nonumber\\
 & & +\frac{C^{KK\ (1)}_s(U_\alpha,{\bar U}_{\bar\alpha})\sqrt{\tau_s}}{{\cal V}\left(\sum_{(m,n)\in{\bf Z}^2/(0,0)}\frac{\frac{(\tau-{\bar\tau})}{2i}}{|m+n\tau|^2}\right)} + \frac{C^{KK\ (1)}_b(U_\alpha,{\bar U}_{\bar\alpha})\sqrt{\tau_b}}{{\cal V}\left(\sum_{(m,n)\in{\bf Z}^2/(0,0)}\frac{\frac{(\tau-{\bar\tau})}{2i}}{|m+n\tau|^2}\right)}\nonumber\\
 & & +\frac{C^{KK\ (2)}_s(U_\alpha,{\bar U}_{\bar\alpha})}{\left(\sum_{(m,n)\in{\bf Z}^2/(0,0)}\frac{\left(\frac{\tau-{\bar\tau}}{2i}\right)^2}{|m+n\tau|^4}\right)}\frac{\partial^2K_{\rm tree}}{\partial\tau_s^2} + \frac{C^{KK\ (2)}_b(U_\alpha,{\bar U}_{\bar\alpha})}{\left(\sum_{(m,n)\in{\bf Z}^2/(0,0)}\frac{\left(\frac{\tau-{\bar\tau})}{2i}\right)^2}{|m+n\tau|^4}\right)}\frac{\partial^2K_{\rm tree}}{\partial\tau_b^2}.
\end{eqnarray}
In (\ref{eq:nonpert81}), the first line and $-2\ ln({\cal V})$ are the tree-level contributions, the second (excluding the volume factor in the argument of the logarithm) and third lines are the perturbative and non-perturbative $\alpha^\prime$ corrections, the fourth line is the 1-loop contribution and the last line is the two-loop contribution; $\tau_s$ is the volume of the ``small" divisor and $\tau_b$ is the volume of the ``big" divisor. The loop-contributions arise from KK modes corresponding to closed string or 1-loop open-string exchange between $D3$- and $D7$-(or $O7$-planes)branes wrapped around the ``s" and ``b" divisors - note that the two divisors do not intersect (See \cite{Curio+Spillner}) implying that there is no contribution from winding modes corresponding to strings winding non-contractible 1-cycles in the intersection locus corresponding to stacks of intersecting $D7$-branes wrapped around the ``s" and ``b" divisors.

Based on (\ref{eq:nonpert81}), the inverse metric (not been careful as regards numerical factors in the numerators and denominators)is given by (dropping 2-loop contributions as they are sub-dominant as compared to the 1-loop contributions):
\begin{eqnarray}
\label{eq:nonpert151}
& & {\cal G}^{-1}=\left(\begin{array}{cccc}
{\cal G}^{\rho_s{\bar\rho_s}} & {\cal G}^{\rho_s{\bar\rho_b}} & {\cal G}^{\rho_s{\bar G^1}} & 0 \\
\overline{{\cal G}^{\rho_s{\bar\rho_b}}} & {\cal G}^{\rho_b{\bar\rho_b}} & {\cal G}^{\rho_b{\bar G^1}} & 0 \\
\overline{{\cal G}^{\rho_s{\bar G^1}}} & \overline{{\cal G}^{\rho_b{\bar G^1}}} & {\cal G}^{G^1{\bar G^1}}
&  {\cal G}^{G^1{\bar G}^2}\\
0 & 0 & {\cal G}^{G^2{\bar G}^1} & {\cal G}^{G^2{\bar G}^2}
\end{array}\right)=\nonumber\\
& & \hskip -0.83in\left(\begin{array}{cccc}
\frac{\frac{\tau-{\bar\tau}}{2i}{\cal Y}(ln {\cal Y})^{\frac{3}{2}}}{\frac{ln {\cal Y}}{ \left(\frac{\tau-{\bar\tau}}{2i}\right)}+\frac{C^{KK\ (1)}_s}{{\cal T}}} & \frac{\frac{{\cal Y}^{\frac{2}{3}}(ln {\cal Y})^2}{\left(\frac{\tau-{\bar\tau}}{2i}\right)}+\frac{C^{KK\ (1)}_s {\cal Y}^{\frac{2}{3}} ln {\cal Y}}{{\cal T}}}{ln {\cal Y} \left(\frac{\tau-{\bar\tau}}{2i}\right)^{-1}+\frac{C^{KK\ (1)}_s}{{\cal T}}} & \frac{i{\cal Z}{\cal Y}^{-1}}{\frac{\tau-{\bar\tau}}{2i}}\frac{\left(-\left(\frac{\tau-{\bar\tau}}{2i}\right)(ln {\cal Y})^2+\frac{C^{KK\ (1)}_s
 ln {\cal Y}}{{\cal T}}\right)}{\frac{\chi_1ln {\cal Y}} {\left(\frac{\tau-{\bar\tau}}{2i}\right)}+\frac{C^{KK\ (1)}_s}{{\cal T}}} & \!\!\!\!\!\!0\\
\frac{\frac{{\cal Y}^{\frac{2}{3}}(ln {\cal Y})^2}{\left(\frac{\tau-{\bar\tau}}{2i}\right)}+\frac{C^{KK\ (1)}_s {\cal Y}^{\frac{2}{3}} ln {\cal Y}}{{\cal T}}}{\frac{ln {\cal Y}} {\left(\frac{\tau-{\bar\tau}}{2i}\right)}+\frac{C^{KK\ (1)}_s}{{\cal T}}} &
\frac{{\cal Y}^{\frac{4}{3}}\left(\frac{ln {\cal Y}}{\left(\frac{\tau-{\bar\tau}}{2i}\right)}
- \frac{C^{KK\ (1)}_s}{{\cal T}}\right)}{\frac{ln {\cal Y}}{\left(\frac{\tau-{\bar\tau}}{2i}\right)}-\frac{C^{KK\ (1)}_s}{{\cal T}}} & \frac{i{\cal Z}{\cal Y}^{-\frac{1}{3}}\left(\frac{\tau-{\bar\tau}}{2i}\right)}{k_1{\cal X}_1}\frac{\left(ln {\cal Y}-\left(\frac{\tau-{\bar\tau}}{2i}\right)\frac{C^{KK\ (1)}_s
 }{{\cal T}}\right)}{\frac{ln {\cal Y}}{\left(\frac{\tau-{\bar\tau}}{2i}\right)}-\frac{C^{KK\ (1)}_s}{{\cal T}}} &\!\!\!\!\!\! 0 \\
 \frac{-i{\cal Z}{\cal Y}^{-1}}{\frac{\tau-{\bar\tau}}{2i}}\frac{\left(-\left(\frac{\tau-{\bar\tau}}{2i}\right)(ln {\cal Y})^2+\frac{C^{KK\ (1)}_s
 ln {\cal Y}}{{\cal T}}\right)}{\frac{\chi_1ln {\cal Y}} {\left(\frac{\tau-{\bar\tau}}{2i}\right)}+\frac{C^{KK\ (1)}_s}{{\cal T}}} &
 \frac{-i{\cal Z}{\cal Y}^{-\frac{1}{3}}\left(\frac{\tau-{\bar\tau}}{2i}\right)}{k_1{\cal X}_1}\frac{\left(ln {\cal Y}-\left(\frac{\tau-{\bar\tau}}{2i}\right)\frac{C^{KK\ (1)}_s
 }{{\cal T}}\right)}{\frac{ln {\cal Y}}{\left(\frac{\tau-{\bar\tau}}{2i}\right)}-\frac{C^{KK\ (1)}_s}{{\cal T}}} & \frac{1}{(k_1^2-k_2^2)\chi_1}\frac{\left(-\frac{ln {\cal Y}}{\left(\frac{\tau-{\bar\tau}}{2i}\right)}
 + \frac{C^{KK\ (1)}_s}{{\cal T}}\right)}{\frac{ln {\cal Y}}{\left(\frac{\tau-{\bar\tau}}{2i}\right)}-\frac{C^{KK\ (1)}_s}{{\cal T}}} & \frac{k_2}{(k_1k_2^2-k_1^3)\chi_1} \\
 & & & \\
 0 & 0 & \frac{k_2}{(k_1k_2^2-k_2^3)\chi_1} & \!\!\!\!\!\!\frac{1}{\chi_1(k_1^2-k_2^2)}\\
 \end{array}\right)\nonumber\\
 & &
\end{eqnarray}
where
\begin{equation}
\label{eq:defs}
{\cal T}\equiv \sum_{(m,n)\in{\bf Z}^2/(0,0)}\frac{\frac{(\tau-{\bar\tau})}{2i}}{|m+n\tau|^2}.
\end{equation}
{\it It becomes evident from (\ref{eq:nonpert81}) and (\ref{eq:nonpert151}) that loop corrections are sub-dominant as compared to the perturbative and non-perturbative $\alpha^\prime$ corrections.} One of the consequences of inclusion of perturbative $\alpha^\prime$-corrections is that the ${\cal N}=1$ potential receives a contribution of the type $\frac{\chi(CY_3)|W_{cs}+W_{np}|^2}{{\cal V}^3}$ ($cs\equiv$ complex structure,
$np\equiv$ non-perturbative) \cite{BBHL}. But, in the approximation that $W_{c.s.}<<1$ (See \cite{KKLT}) and further assuming that $W_{c.s.}<W_{np}$, this contribution (given by $\frac{1}{{\cal V}^{3+2n^s}}$) is sub-dominant as compared to the contribution from the $D1$-brane and $D3$-brane instanton superpotential, e.g.,
$G^{\rho^s{\bar\rho}^s}\partial_{\rho^s}W_{np}{\bar\partial}_{\bar\rho^s}{\bar W}_{np}+c.c.\sim\frac{\sqrt{ln{\cal V}}}{{\cal V}^{2n^s-1}}$ for $D3-$brane instanton number $n^s>1$. To be a bit more detailed conceptually,
using (\ref{eq:nonpert81}) and (\ref{eq:nonpert15}) and appropriate expression for $W_{np}$, one can show that the ${\cal N}=1$ potential including tree-level (denoted by ``tree"), 1-loop (denoted by ``$g_s$") as well as perturbative (denoted by ``$\chi(CY_3)$") and non-perturbative (denoted by ``$\alpha^\prime_{np}")
\alpha^\prime$ corrections is of the form: $$ e^K G^{\rho^s{\bar\rho}^s}|_{({\rm tree},\ g_s,\chi(CY_3),\ \alpha^\prime_{np})}\left|{\bar\partial}_{\bar\rho^s}\left({\bar W}_{D1-{\rm instanton}}({\bar\tau},{\bar G}^a){\bar W}_{D3-{\rm instanton}}({\bar\tau},{\bar G}^a,{\bar\rho}^s,{\bar\rho}^b)\right)\right|^2.$$ In the LVS limit for the ``Swiss cheese" considered in our paper,
for extremization calculations, one can equivalently consider the following potential:
$$e^KG^{\rho^s{\bar\rho}^s}|_{\rm tree}\left|{\bar\partial}_{\bar\rho^s}\left({\bar W}_{D1-{\rm instanton}}({\bar\tau},{\bar G}^a){\bar W}_{D3-{\rm instanton}}({\bar\tau},{\bar G}^a,{\bar\rho}^s,{\bar\rho}^b) \right)\right|^2.$$

Having extremized the superpotential w.r.t. the complex structure moduli and the axion-dilaton
modulus, the ${\cal N}=1$ potential
will be given by:
\begin{eqnarray}
\label{eq:nonpert17}
& & V = e^K\Biggl[\sum_{A,B=\rho_\alpha,G^a}\Biggl\{({\cal G}^{-1})^{A{\bar B}}\partial_A W_{np}{\bar\partial}_{\bar B}{\bar W_{np}}
+ \left(({\cal G}^{-1})^{A{\bar B}}(\partial_A K){\bar\partial}_{\bar B}{\bar W_{np}} + c.c.\right)\Biggr\} \nonumber\\
& & + \left(\sum_{A,B=\rho_\alpha,G^a}({\cal G}^{-1})^{A{\bar B}}\partial_A K{\bar\partial}_{\bar B}K - 3\right)|W|^2 + \sum_{\alpha,{\bar\beta}\in{\rm c.s.}}({\cal G}^{-1})^{\alpha{\bar\beta}}\partial_{\alpha} K_{c.s.}{\bar\partial}_{\bar\beta}K_{c.s.}|W_{np}|^2
\Biggr],
\end{eqnarray}
where the total superpotential $W$ is the sum of the complex structure moduli Gukov-Vafa-Witten superpotential
and the non-perturbative superpotential $W_{np}$ arising because of instantons (obtained by wrapping of
$D3$-branes around the divisors with complexified volumes $\tau_s$ and $\tau_b$).

To summarize the result of section 4 of \cite{SwissCheeseissues}, one gets the following potential:
\begin{eqnarray}
\label{eq:nonpert21}
& & V\sim\frac{{\cal Y}\sqrt{ln {\cal V}}}{{\cal V}^{2n^s+2}}e^{-2\phi}(n^s)^2\frac{\left(\sum_{m^a}e^{-\frac{m^2}{2g_s} + \frac{m_ab^a n^s}{g_s} + \frac{n^s\kappa_{1ab}b^ab^b}{2g_s}}\right)^2}{\left|f(\eta(\tau))\right|^2}
\nonumber\\
& & +\frac{ln {\cal V}}{{\cal V}^{n^s+2}}\left(\frac{\theta_{n^s}({\bar\tau},{\bar G})}{f(\eta({\bar\tau}))}
\right)e^{-in^s(-\tilde{\rho_1}+\frac{1}{2}\kappa_{1ab}
\frac{{\bar\tau}G^a-\tau{\bar G}^a}{({\bar\tau}-\tau)}\frac{(G^b-{\bar G}^b)}{({\bar\tau}-\tau)} -
\frac{1}{2}\kappa_{1ab}\frac{G^a(G^b-{\bar G}^b)}{(\tau-{\bar\tau})})}+c.c.\nonumber\\
& & +
\frac{|W|^2}{{\cal V}^3}\left(\frac{3k_2^2+k_1^2}{k_1^2-k_2^2}\right)
\frac{\left|\sum_c\sum_{n,m\in{\bf Z}^2/(0,0)}e^{-\frac{3\phi}{2}}A_{n,m,n_{k^c}}(\tau) sin(nk.b+mk.c)\right|^2}
{\sum_{c^\prime}\sum_{m^\prime,n^\prime\in{\bf Z}^2/(0,0)} e^{-\frac{3\phi}{2}}|n+m\tau|^3
|A_{n^\prime,m^\prime,n_{k^{c^{\prime}}}}(\tau)|^2 cos(n^\prime k.b+m^\prime k.c)}+\frac{\xi|W|^2}{{\cal V}^3}.
\nonumber\\
& &
\end{eqnarray}
On comparing (\ref{eq:nonpert21}) with the analysis of \cite{Balaetal2}, one sees that for generic values of
the moduli $\rho_\alpha, G^a, k^{1,2}$ and ${\cal O}(1)\ W_{c.s.}$, and $n^s$(the $D3$-brane instanton quantum number)=1, analogous to \cite{Balaetal2}, the second term
dominates; the third term is a new term. However, as in KKLT scenarios (See \cite{KKLT}), $W_{c.s.}<<1$; we would henceforth assume that the fluxes and complex structure moduli have been so fine tuned/fixed that $W\sim W_{n.p.}$. We assume that the fundamental-domain-valued $b^a$'s satisfy: $\frac{|b^a|}{\pi}<1$\footnote{If one puts in appropriate powers of the Planck mass $M_p$, $\frac{|b^a|}{\pi}<1$ is equivalent to $|b^a|<M_p$, i.e., NS-NS axions are sub-Planckian in units of $\pi M_p$.}. This implies that for $n^s>1$, the first term in (\ref{eq:nonpert21}) - $|\partial_{\rho^1}W_{np}|^2$ - a positive definite term and denoted henceforth by $V_I$, is the most dominant. Hence, if a minimum exists, it will be positive. As shown in \cite{SwissCheeseissues}, the potential can be extremized along the locus:
\begin{equation}
\label{eq:ext_locus}
mk.c + nk.b = N_{(m,n;,k^a)}\pi
\end{equation}
and very large values of the $D1$-instanton quantum numbers $m^a$.
As shown in section {\bf 3}, it turns out that the locus $nk.b + mk.c = N\pi$ for $|b^a|<\pi$ and $|c^a|<\pi$ corresponds to a flat saddle point with the NS-NS axions providing a flat direction.

Analogous to \cite{Balaetal2}, for all directions in the moduli space with ${\cal O}(1)$ $W_{c.s.}$ and away from $D_iW_{cs}=D_\tau W=0=\partial_{c^a}V=\partial_{b^a}V=0$, the ${\cal O}(\frac{1}{{\cal V}^2})$ contribution
of $\sum_{\alpha,{\bar\beta}\in{c.s.}}(G^{-1})^{\alpha{\bar\beta}}D_\alpha W_{cs}{\bar D}_{\bar\beta}{\bar W}_{cs}$  dominates over (\ref{eq:nonpert21}),
ensuring that that there must exist a minimum, and given the positive definiteness of the potential $V_I$, this will be a dS minimum. There has been no need to add any $\overline{D3}$-branes as in KKLT to generate a dS vacuum. Also, interestingly, as we show in section {\bf 3}, the condition
$nk.b + mk.c = N \pi$ gurantees that the slow roll parameters ``$\epsilon$" and ``$\eta$" are much smaller than one for slow roll inflation beginning from the saddle point and proceeding along an NS-NS axionic flat direction towards the nearest dS minimum.

The arguments related to the life-time of the dS minimum in the literature estimate the lifetime to be
$\sim e^{\frac{2\pi^2}{V_0}}$ where the minimum value of the potential, $V_0\sim \frac{\sqrt{ln {\cal V}}}{{\cal V}^N}$ for $N\geq5$. The lifetime, hence, can be made arbitrarily large as ${\cal V}$ is increased.

\section{Axionic Slow Roll Inflation}

In this section, we discuss the possibility of getting slow roll inflation along a flat direction provided by the NS-NS axions starting from a saddle point and proceeding to the nearest dS minimum. In what follows, we will assume that the volume moduli for the small and big divisors and the axion-dilaton modulus have been stabilized. All calculations henceforth will be in the axionic sector - $\partial_a$ will imply $\partial_{G^a}$ in the following.

We need now to evaluate the slow-roll inflation parameters (in $M_p=1$ units)
$\epsilon\equiv\frac{{\cal G}^{ij}\partial_iV\partial_jV}{2V^2},\ \eta\equiv$ the most negative eigenvalue of the matrix $N^i_{\ j}\equiv\frac{{\cal G}^{ik}\left(\partial_k\partial_jV - \Gamma^l_{jk}\partial_lV\right)}{V}$. In terms of the real axions,
\begin{equation}
\label{eq:realN}
N=\left(\begin{array}{cccc}
N^{c^1}_{\ c^1} & N^{c^1}_{\ c^2} & N^{c^1}_{\ b^1} & N^{c^1}_{b^2} \\
N^{c^2}_{\ c^1} & N^{c^2}_{\ c^2} & N^{c^2}_{\ b^1} & N^{c^2}_{b^2} \\
N^{b^1}_{\ c^1} & N^{b^1}_{\ c^2} & N^{b^1}_{\ b^1} & N^{b^1}_{b^2} \\
N^{b^2}_{\ c^1} & N^{b^2}_{\ c^2} & N^{b^2}_{\ b^1} & N^{b^2}_{b^2} \\
\end{array}\right).
\end{equation}
In terms of the complex $G^{1,2}$ and ${\bar G}^{{\bar 1},{\bar 2}}$,
\begin{eqnarray}
\label{eq:compN}
& & N^{c^1}_{\ c^1}=\frac{{\bar\tau}}{{\bar\tau}-\tau}N^{G^1}_{\ G^1}
-\frac{\tau}{{\bar\tau}-\tau}N^{{\bar G}^{\bar 1}}_{\ G^1}
+\frac{{\bar\tau}}{{\bar\tau}-\tau}N^{G^1}_{\ {\bar G}^{\bar 1}}
-\frac{\tau}{{\bar\tau}-\tau}N^{{\bar G}^{\bar 1}}_{\ G^1},\nonumber\\
& & N^{c^1}_{\ c^2}=\frac{{\bar\tau}}{{\bar\tau}-\tau}N^{G^1}_{\ G^2}
+\frac{\tau}{{\bar\tau}-\tau}N^{G^1}_{\ {\bar G}^{\bar 2}}
-\frac{\tau}{{\bar\tau}-\tau}N^{{\bar G}^{\bar 1}}_{\ G^2}
-\frac{\tau}{{\bar\tau}-\tau}N^{{\bar G}^{\bar 1}}_{\ {\bar G}^{\bar 2}},\nonumber\\
& & N^{c^1}_{\ b^1}=-\frac{|\tau|^2}{{\bar\tau}-\tau}N^{G^1}_{\ G^1}
+\frac{\tau^2}{{\bar\tau}-\tau}N^{{\bar G}^{\bar 1}}_{\ G^1}
-\frac{{\bar\tau}^2}{{\bar\tau}-\tau}N^{G^1}_{\ {\bar G}^{\bar 1}}
+\frac{|\tau|^2}{{\bar\tau}-\tau}N^{{\bar G}^{\bar 1}}_{\ {\bar G}^1},\nonumber\\
& & N^{c^1}_{\ b^2}=-\frac{|\tau|^2}{{\bar\tau}-\tau}N^{G^1}_{\ G^2}
-\frac{|\tau|^2}{{\bar\tau}-\tau}N^{{\bar G}^{\bar 1}}_{\ G^2}
+\frac{\tau}{{\bar\tau}-\tau}N^{G^1}_{\ {\bar G}^{\bar 2}}
+\frac{|\tau|^2}{{\bar\tau}-\tau}N^{{\bar G}^{\bar 1}}_{\ {\bar G}^{\bar 2}},\ {\rm etc}.
\end{eqnarray}

The first derivative of the potential is given by:
\begin{equation}
\label{eq:dV}
\partial_aV|_{D_{c.s.}W=D_\tau W=0}=(\partial_a K)V+e^K\biggl[{\cal G}^{\rho_s{\bar\rho}_s}((\partial_a \partial_{\rho_s}W_{np} {\bar\partial}_{\bar\rho_s}){\bar W}_{np}+\partial_{\rho_s}W_{np}\partial_a{\bar\partial_{\bar\rho_s}}{\bar W}_{np})+\partial_a{\cal G}^{\rho_s{\bar\rho}_s}\partial_{\rho_s}W_{np}{\bar\partial}_{\bar\rho_s}{\bar W}_{np}\biggr].
\end{equation}
The most dominant terms in (\ref{eq:dV}) of ${\cal O}(\frac{\sqrt{ln {\cal V}}}{{\cal V}^{2n^s+1}})$ that potentially violate the requirement ``$\epsilon<<1$" are of the type:
\begin{itemize}
\item
e.g. $e^K(\partial_a{\cal G}^{\rho_s{\bar\rho}_s})(\partial_bW_{np}){\bar\partial}_{\bar c}{\bar W}_{np}$, is proportional to $\partial_a\chi_2$, which at the locus
(\ref{eq:ext_locus}), vanishes;

\item
e.g. $e^K {\cal G}^{\rho_s{\bar\rho}_s}\partial_a\partial_b W_{np}{\bar\partial}_{\bar c}{\bar W}_{np}$: the contribution to $\epsilon$ will be
$\frac{{\cal V}}{\sum_{\beta\in H_2}(n^0_\beta)^2}$. Now, it turns out that the genus-0 Gopakumar-Vafa integer invariants $n^0_\beta$'s for compact Calabi-Yau's of a projective variety in weighted complex projective spaces for appropriate degree of the holomorphic curve, can be as large as $10^{20}$ and even higher \cite{Klemm_GV} thereby guaranteeing that the said contribution to $\epsilon$ respects the slow roll inflation requirement.
\end{itemize}
 One can hence show from (\ref{eq:dV}) that along (\ref{eq:ext_locus}), $\epsilon<<1$ is always satisfied.

To evaluate $N^a_{\  b}$ and the Hessian, one needs to evaluate the second derivatives of the potential and components of the affine connection. In this regard, one needs to evaluate, e.g.:
\begin{eqnarray}
\label{eq:ddV}
& & {\bar\partial}_{\bar d}\partial_aV=({\bar\partial}_{\bar d}\partial_aK)V+\partial_aK{\bar\partial}_{\bar d}V\nonumber\\
& & +e^K\biggr[{\bar\partial}_{\bar d}\partial_a{\cal G}^{\rho_s{\bar\rho_s}}\partial_{\rho_s}W_{np}{\bar\partial}_{\bar\rho_s}{\bar W}_{np}
+ \partial_a{\cal G}^{\rho_s{\bar\rho_s}}{\bar\partial}_{\bar d}\left(\partial_{\rho_s}W_{np}{\bar\partial}_{\bar\rho_s}{\bar W}_{np}\right)
+ {\bar\partial}_{\bar d}{\cal G}^{\rho_s{\bar\rho_s}}\partial_a\left(\partial_{\rho_s}W_{np}{\bar\partial}_{\bar\rho_s}{\bar W}_{np}\right)\nonumber\\
 & & + {\cal G}^{\rho_s{\bar\rho_s}}\partial_a{\bar\partial}_{\bar d}\left(\partial_{\rho_s}W_{np}{\bar\partial}_{\bar\rho_s}{\bar W}_{np}\right)\biggr].
\end{eqnarray}
One can show that at (\ref{eq:ext_locus}), the most dominant term (and hence the most dominant contribution to $\eta$) in (\ref{eq:ddV}) comes from
$e^K{\cal G}^{\rho_s{\bar\rho}_s}\partial_b\partial_{\rho_s}W_{np}{\bar\partial}_{\bar c}{\bar\partial_{\bar\rho_s}}{\bar W}_{np}$, proportional to:
\begin{equation}
\label{eq:dom_eta_contr}
N^{\bar a}_{\ b}\ni \frac{(n^s)^2{\cal V}}{\sum_{\beta\in H_2} (n^0_\beta)^2}.
\end{equation}
Now, the large values of the genus-0 Gopakumar-Vafa invariants again nullifies this contribution to $\eta$.

Now, the affine connection components, in the LVS limit, are given by:
\begin{equation}
\label{eq:affine}
\Gamma^a_{bc}={\cal G}^{a{\bar d}}\partial_b{\cal G}_{c{\bar d}}\sim\left[\left(\frac{{\bar\tau}}{{\bar\tau}-\tau}\right)\partial_{c^a}
+\left(\frac{1}{{\bar\tau}-\tau}\right)\partial_{b^a}\right]{\cal X}_1\equiv {\cal O}({\cal V}^0),
\end{equation}
implying that
\begin{equation}
\label{eq:affineContrN}
N^{\bar a}_{\ b}\ni\frac{{\cal G}^{c{\bar a}}\Gamma^d_{cb}\partial_dV}{V}\sim\frac{{\cal V}
\sum_{m,n\in{\bf Z}^2/(0,0)}
\frac{({\bar\tau}-\tau)^{\frac{3}{2}}}{(2i)^{\frac{3}{2}}|m+n\tau|^3} sin(nk.b + mk.c)\frac{\sqrt{ln {\cal V}}}{{\cal V}^{1 + 2n^s}}}{\sum_{\beta\in H_2^-(CY_3,{\bf Z})} (n^0_\beta)^2\frac{\sqrt{ln{\cal V}}}{{\cal V}^{1 + 2n^s}}}.
\end{equation}
We thus see that in the LVS limit and because of the large genus-0 Gopakumar-Vafa invariants, this contribution is nullified - note that near the locus (\ref{eq:ext_locus}), the contribution is further damped. Thus the ``$\eta$ problem" of \cite{KKLMMT} is solved.

We will  show that one gets a saddle point at $\{(b^a,c^a)|nk.b + mk.c=N_{(m,n;k^a)}\pi\}$ and the NS-NS axions provide a flat direction. We will work out the slow-roll inflation direction along which inflation proceeds between the saddle point and the minimum. Now,
the Hessian or the mass matrix ${\cal M}$ of fluctuations is defined as:
\begin{equation}
\label{eq:Hessian}
{\cal M}=\left(\begin{array}{cc}
2{\rm Re}\left(\partial_a{\partial_{\bar b}}V + \partial_a\partial_bV\right) & -2{\rm Im}\left(\partial_a{\bar\partial_{\bar b}}V+\partial_a\partial_bV\right)\\
-2{\rm Im}\left(\partial_a{\bar\partial_{\bar b}}V-\partial_a\partial_bV\right) &
2{\rm Re}\left(\partial_a{\partial_{\bar b}}V - \partial_a\partial_bV\right)\\
\end{array}\right).
\end{equation}
An eigenvector of the Hessian is to be understood to denote the following fluctuation direction:
\begin{equation}
\label{eq:evectordirec}
\left(\begin{array}{c}
\delta c^1 - A \delta b^1 \\
\delta c^2 - A \delta b^2 \\
-\frac{1}{g_s}\delta b^1 \\
-\frac{1}{g_s}\delta b^2
\end{array}\right).
\end{equation}

One can show that near $nk.b+mk.c=N\pi$ and $b^a\sim-\frac{m^a}{\kappa}\sim\frac{N\pi}{nk^a}$, assuming that $\frac{nk.m}{\pi\kappa}\in{\bf Z}$:
\begin{eqnarray}
\label{eq:GVvolmore1}
& & \partial_a\partial_bV=\Lambda_1{\bar\tau}^2n^2k_ak_b + \Lambda_1{\bar\tau}nmk_ak_b +\Lambda_2|\kappa_{1ab}|,\nonumber\\
& & \partial_a{\bar\partial_{\bar b}}V=-\Lambda_1|\tau|^2n^2k_ak_b-\Lambda_1{\bar\tau}nmk_ak_b-\Lambda_2|\kappa_{1ab}|,
\end{eqnarray}
where
\begin{eqnarray}
\label{eq:Lambdas}
& & \Lambda_1\equiv\frac{4}{|\tau-{\bar\tau}|^2}\frac{\sqrt{ln{\cal V}}}{\cal V}\sum_{\beta\in H_2^-}\frac{n^0_\beta}{\cal V}\sum_{(m,n)\in{\bf Z}^2/(0,0)}\frac{\left(\frac{\tau - {\bar\tau}}{2i}\right)^{\frac{3}{2}}}{|m+n\tau|^3}\frac{\left(\sum_{m^a}e^{-\frac{m^2}{2g_s} + \frac{m_ab^a n^1}{g_s} + \frac{n^1\kappa_{1ab}b^ab^b}{2g_s}}\right)^2}{\left|f(\eta(\tau))\right|^2}\sim{\cal O}(g_s^2),
\nonumber\\
& & \Lambda_2\equiv\frac{2}{|\tau-{\bar\tau}|^2}\frac{\sqrt{ln{\cal V}}}{\cal V}\frac{\left(\sum_{m^a}e^{-\frac{m^2}{2g_s} + \frac{m_ab^a n^1}{g_s} + \frac{n^1\kappa_{1ab}b^ab^b}{2g_s}}\right)^2}{\left|f(\eta(\tau))\right|^2}
\sum_{m^a,\ {\rm no\ sum\ over\ }a}e^{-\frac{m^2}{2g_s} + \frac{m_ab^a n^1}{g_s} + \frac{n^1\kappa_{1ab}b^ab^b}{2g_s}}\sim{\cal O}(g_s^2).\nonumber\\
& &
\end{eqnarray}
In the limit $A>>1$, one gets the Hessian:
\begin{equation}
\left(\begin{array}{cccc}
-\frac{2}{g_s^2}\Lambda_1n^2k_1^2 & -\frac{2}{g_s^2}n^2k_1k_2 & \frac{2A}{g_s}\Lambda_1n^2k_1^2 &
\frac{2A}{g_s}\Lambda_1n^2k_1k_2\\
-\frac{2}{g_s^2}\Lambda_1n^2k_1k_2 & -\frac{2}{g_s^2}n^2k_2^2 & \frac{2A}{g_s}\Lambda_1n^2k_1k_2 &
\frac{2A}{g_s}\Lambda_1n^2k_2^2\\
\frac{2A}{g_s^2}\Lambda_1n^2k_1^2 & \frac{2A}{g_s^2}n^2k_1k_2 & 2A^2\Lambda_1n^2k_1^2 - |{\cal X}|&
2A^2\Lambda_1n^2k_1k_2\\
\frac{2A}{g_s^2}\Lambda_1n^2k_1k_2 & \frac{2A}{g_s^2}n^2k_2^2 & 2A^2\Lambda_1n^2k_1k_2 &
2A^2\Lambda_1n^2k_2^2 - |{\cal X}|
\end{array}\right),
\end{equation}
where ${\cal X}\equiv 2\Lambda_2|\kappa_{1ab}|\sim{\cal O}(g_s^2)$.
The eigenvalues are given by:
$$\{0,-|{\cal X}|,\frac{2 A^2 {k_1}^2 {\Lambda_1} n^2 g_s^3+2 A^2 {k_2}^2 {\Lambda_1} n^2 g_s^3-|{\cal X}| g_s^3-2 {k_1}^2 {\Lambda_1} n^2 g_s-2 {k_2}^2 {\Lambda_1} n^2 g+\sqrt{\cal Z}}{2 g_s^3},$$\\
$$   -\frac{-2 A^2 {k_1}^2 {\Lambda_1}
   n^2 g_s^3-2 A^2 {k_2}^2 {\Lambda_1} n^2 g_s^3+|{\cal X}| g_s^3+2 {k_1}^2 {\Lambda_1} n^2 g_s+2 {k_2}^2 {\Lambda_1} n^2 g_s+\sqrt{\cal Z}}{2 g_s^3}\},$$where
   $${\cal Z}\equiv g_s^2
   \biggl(8 g_s \left({k_1}^2+{k_2}^2\right) {\Lambda_1} \left(2 A^2 (g_s+1) \left({k_1}^2+{k_2}^2\right) {\Lambda_1} n^2-g_s |{\cal X}|\right) n^2$$
   $$+\left(-|{\cal X}| g_s^2+2
   \left(A^2 g_s^2-1\right) {k_1}^2 {\Lambda_1} n^2+2 \left(A^2 g_s^2-1\right) {k_2}^2 {\Lambda_1} n^2\right)^2\biggr).$$
The eigenvectors are given by:
\begin{eqnarray}
\label{eq:evectors}
&& \left(
\begin{array}{c}
 -\frac{{k_2}}{{k_1}} \\ 1 \\ 0 \\ 0 \end{array}\right) \nonumber\\
& & \left(\begin{array}{c} 0 \\ 0 \\ -\frac{{k_2}}{{k_1}} \\ 1\end{array}\right) \nonumber\\
& & \left(\begin{array}{c}
 -\frac{{k_1} \left(2 A^2 {k_1}^2 {\Lambda_1} n^2 g_s^3+2 A^2 {k_2}^2 {\Lambda_1} n^2 g_s^3-|{\cal X}| g_s^3+4 A^2 {k_1}^2 {\Lambda_1} n^2 g_s^2+4 A^2 {k_2}^2 {\Lambda_1}
   n^2 g_s^2+2 {k_1}^2 {\Lambda_1} n^2 g_s+2 {k_2}^2 {\Lambda_1} n^2 g_s+\sqrt{\cal Z}\right)}{A g_s {k_2} \left(-2 A^2 {k_1}^2 {\Lambda_1} n^2 g_s^3-2 A^2 {k_2}^2 {\Lambda_1} n^2 g_s^3+|{\cal X}| g_s^3+2
   {k_1}^2 {\Lambda_1} n^2 g_s+2 {k_2}^2 {\Lambda_1} n^2 g_s+\sqrt{\cal Z}\right)} \\
   -\frac{2 A^2 {k_1}^2 {\Lambda_1} n^2 g_s^3+2 A^2 {k_2}^2 {\Lambda_1} n^2 g_s^3-|{\cal X}| g_s^3+4 A^2 {k_1}^2
   {\Lambda_1} n^2 g_s^2+4 A^2 {k_2}^2 {\Lambda_1} n^2 g_s^2+2 {k_1}^2 {\Lambda_1} n^2 g_s+2 {k_2}^2 {\Lambda_1} n^2 g_s+\sqrt{\cal Z}}{A g_s \left(-2 A^2 {k_1}^2 {\Lambda_1} n^2 g_s^3-2 A^2
   {k_2}^2 {\Lambda_1} n^2 g_s^3+|{\cal X}| g_s^3+2 {k_1}^2 {\Lambda_1} n^2 g_s+2 {k_2}^2 {\Lambda_1} n^2 g_s+\sqrt{\cal Z}\right)}
   \\ \frac{{k_1}}{{k_2}} \\ 1
   \end{array}\right)\nonumber\\
& & \left(\begin{array}{c}
 \frac{{k_1} \left(2 A^2 {k_1}^2 {\Lambda_1} n^2 g_s^3+2 A^2 {k_2}^2 {\Lambda_1} n^2 g_s^3-|{\cal X}| g_s^3+4 A^2 {k_1}^2 {\Lambda_1} n^2 g_s^2+4 A^2 {k_2}^2 {\Lambda_1}
   n^2 g_s^2+2 {k_1}^2 {\Lambda_1} n^2 g_s+2 {k_2}^2 {\Lambda_1} n^2 g_s-\sqrt{\cal Z}\right)}{A g_s {k_2} \left(2 A^2 {k_1}^2 {\Lambda_1} n^2 g_s^3+2 A^2 {k_2}^2 {\Lambda_1} n^2 g_s^3-|{\cal X}| g_s^3-2 {k_1}^2
   {\Lambda_1} n^2 g_s-2 {k_2}^2 {\Lambda_1} n^2 g_s+\sqrt{\cal Z}\right)} \\
   \frac{2 A^2 {k_1}^2 {\Lambda_1} n^2 g_s^3+2 A^2 {k_2}^2 {\Lambda_1} n^2 g_s^3-|{\cal X}| g_s^3+4 A^2 {k_1}^2
   {\Lambda_1} n^2 g_s^2+4 A^2 {k_2}^2 {\Lambda_1} n^2 g_s^2+2 {k_1}^2 {\Lambda_1} n^2 g_s+2 {k_2}^2 {\Lambda_1} n^2 g_s-\sqrt{\cal Z}}{A g_s \left(2 A^2 {k_1}^2 {\Lambda_1} n^2 g_s^3+2 A^2
   {k_2}^2 {\Lambda_1} n^2 g_s^3-|{\cal X}| g_s^3-2 {k_1}^2 {\Lambda_1} n^2 g_s-2 {k_2}^2 {\Lambda_1} n^2
   g_s+\sqrt{\cal Z}\right)} \\
  \frac{{k_1}}{{k_2}} \\
    1
\end{array}
\right)
\end{eqnarray}
From the second eigenvector in (\ref{eq:evectors}), one sees that the NS-NS axions provide a flat direction. From the set of eigenvalues, one sees that for $g_s<<1$, the fourth eigenvalue is negative and hence the corresponding fourth eigenvector in (\ref{eq:evectors}) provides the unstable direction. One sees that for $g_s<<1$,
the eigenvectors are insensitive to $|{\cal X}|$. Further, in
the fourth eigenvector in (\ref{eq:evectors}), the top two components are of the type $\frac{{\cal O}(g_s^3)}{{\cal O}(g_s^2)}={\cal O}(g_s)$ and hence negligible as compared to the third and fourth components in the same eigenvector - this justifies taking a linear combination of the NS-NS axions
as flat unstable directions for the slow-roll inflation to proceed. 

The kinetic energy terms for the NS-NS and RR axions can be written as:
\begin{equation}
\label{eq:kinax1}
\left(\begin{array}{cccc}
\partial_\mu c^1 & \partial_\mu c^2 & \partial_\mu b^1 & \partial_\mu b^2
\end{array}\right) {\cal K} \left(\begin{array}{c} \partial^\mu c^1 \\ \partial^\mu c^2 \\ \partial^\mu b^1 \\ \partial^\mu b^2 \end{array}\right),
\end{equation}
where
\begin{equation}
\label{eq:kinax2}
{\cal K}\equiv{\cal X}_1\left(\begin{array}{cccc}
k_1^2 & k_1k_2 & -(\tau+{\bar\tau})k_1^2 & -(\tau+{\bar\tau})k_1k_2 \\
k_1k_2 & k_2^2 & -(\tau+{\bar\tau})k_1k_2 & -(\tau+{\bar\tau})k_1k_2 \\
-(\tau+{\bar\tau})k_1^2 & -(\tau+{\bar\tau})k_1k_2 & |\tau|^2k_1^2 & |\tau|^2k_1k_2 \\
-(\tau+{\bar\tau})k_1k_2 & -(\tau+{\bar\tau})k_2^2 & k_1k_2|\tau|^2 & k_2^2|\tau|^2
\end{array}\right).
\end{equation}
Writing $\tau=A+\frac{i}{g_s}$, the eigenvalues of ${\cal K}$ are given by:
\begin{equation}
\label{eq:diagevs}
{\cal X}_1\left\{ 0,0,\frac{\left( 1 + \left( 1 + A^2 \right) \,g_s^2 +
       {\sqrt{\cal S}} \right) \,
     \left( {{k_1}}^2 + {{k_2}}^2 \right) }{2\,g_s^2},
  \frac{\left( 1 + \left( 1 + A^2 \right) \,g_s^2 - {\sqrt{\cal S}} \right) \,\left( {{k_1}}^2 + {{k_2}}^2 \right) }{2\,g_s^2}\right\}
\end{equation}
where ${\cal S}\equiv 1 + 2\,\left( -1 + A^2 \right) \,g_s^2 + \left( 1 + 14\,A^2 + A^4 \right) \,g_s^4$.
The basis of axionic fields that would diagonalize the kinetic energy terms is given by:
\begin{equation}
\label{eq:diagbasis}
\left(\matrix{ \frac{{k_1}\,\left( {b^2}\,{k_1} - {b^1}\,{k_2} \right) \,
     {\sqrt{1 + \frac{{{k_2}}^2}{{{k_1}}^2}}}}{{{k_1}}^2 + {{k_2}}^2} \cr \frac{{k_1}\,
     \left( {c2}\,{k_1} - {c^1}\,{k_2} \right) \,
     {\sqrt{1 + \frac{{{k_2}}^2}{{{k_1}}^2}}}}{{{k_1}}^2 + {{k_2}}^2} \cr \frac{{k_2}\,
     \Omega_1\,
     \left( {b^1}\,\left( 1 + \left( -1 + A^2 \right) \,g_s^2 +
          {\sqrt{\cal S}} \right) \,{k_1} +
       A^2\,{b^2}\,g_s^2\,{k_2} + {b^2}\,
        \left( 1 - g_s^2 + {\sqrt{\cal S}} \right) \,
        {k_2} - 4\,A\,g_s^2\,\left( {c^1}\,{k_1} + {c2}\,{k_2} \right)  \right) }{4\,
     {\sqrt{2}}\,{\sqrt{\cal S}}\,
     \left( {{k_1}}^2 + {{k_2}}^2 \right) } \cr \frac{{k_2}\,
     \Omega_1\,
     \left( {b^1}\,\left( -1 - \left( -1 + A^2 \right) \,g_s^2 +
          {\sqrt{\cal S}} \right) \,{k_1} -
       A^2\,{b^2}\,g_s^2\,{k_2} + {b^2}\,
        \left( -1 + g_s^2 + {\sqrt{\cal S}} \right) \,
        {k_2} + 4\,A\,g_s^2\,\left( {c^1}\,{k_1} + {c2}\,{k_2} \right)  \right) }{4\,
     {\sqrt{2}}\,{\sqrt{\cal S}}\,
     \left( {{k_1}}^2 + {{k_2}}^2 \right) } \cr  }\right),
\end{equation}
where $\Omega_1\equiv {\sqrt{-\left( \frac{\left( -1 - \left( 1 + 14\,A^2 + A^4 \right) \,g_s^4 +
               {\sqrt{\cal S}} +
               \left( -1 + A^2 \right) \,g_s^2\,\left( -2 +
                  {\sqrt{\cal S}} \right)  \right) \,
             \left( {{k_1}}^2 + {{k_2}}^2 \right) }{A^2\,g_s^4\,{{k_2}}^2} \right) }}.$
This tells us that in the $g_s<<1$ limit, there are two NS-NS axionic basis fields in terms of which the axionic kinetic terms are diagonal -
${\cal B}^1\equiv\frac{\left( b^2k_1 - b^1k_2 \right)}
     {\sqrt{k_1^2 + k_2^2}}$, and
     ${\cal B}^2\equiv\frac{1}{2g_s\sqrt{2k_2^2(k_1^2+k_2^2)}}(b^1k_1 + b^2k_2)$.
By solving for $b^1$ and $b^2$ in terms of
${\cal B}^1$ and ${\cal B}^2$, and plugging into the mass term, one finds that the mass term for $B^2$ and not $B^1$, becomes proportional to $g_s^2(B^2)^2$ - given that the inflaton must be lighter than its non-inflatonic partner, one concludes that $\frac{1}{2g_s\sqrt{2k_2^2(k_1^2+k_2^2)}}(b^1k_1 + b^2k_2)$ must be identified with the inflaton. We need to consider a situation wherein one can not completely disregard $\frac{n^0_\beta}{\cal V}$ as compared to unity - this ratio could be smaller than unity but not negligible. This is because the eigenvalues and hence the eigenvectors of the Hessian are more sensitive to this ratio than the term $|{\cal X}|$ that one gets by assuming $\frac{n^0_\beta}{\cal V}<<1$ - in the latter case, one can show that one can not get a nearly flat unstable direction for slow roll to proceed.

\section{Discussion}

In this note, we have generalized the idea in \cite{SwissCheeseissues} of obtaining a dS minimum (using perturbative and non-perturbative corrections to the K\"{a}hler potential and instanton corrections to the superpotential) without the addition of $\overline{D3}$-branes by including the one- and two- loop corrections to the K\"{a}hler potential and showing that two-loop contributions are subdominant w.r.t. one-loop corrections and the one-loop corrections are sub-dominant w.r.t. the perturbative and non-perturbative $\alpha^\prime$ corrections in the {\it LVS} limits.  Assuming the NS-NS and RR axions $b^a, c^a$'s to lie in the fundamental-domain and to satisfy: $\frac{|b^a|}{\pi}<1,\ \frac{|c^a|}{\pi}<1$,  the $D3$-brane instanton number $n^s$ associated with the ``small divisor" to be much larger than the $D1$-instanton numbers $m_{D1}^a$'s, one gets a flat direction provided by the NS-NS axions for slow roll inflation to occur starting from a saddle point and proceeding to the nearest dS minimum. After a detailed calculation we find that for $\epsilon << 1$ in the {\it LVS} limit all along the slow roll. The ``eta problem" gets solved at and away from the saddle point locus for some quantized values  of a linear combination of the NS-NS and RR axions; the slow-roll flat direction is provided by the NS-NS axions. A linear combination of the axions gets identified with the inflaton. Thus in a
nutshell, we have shown the possibility of axionic slow roll inflation in the large volume limit of type IIB compactifications on orientifolds of Swiss Cheese Calabi-Yau's. As a linear combination of the NS-NS axions corresponds to the inflaton in our work, this corresponds to a discretized expansion rate and analogous to \cite{discreteinflation}
may correspond to a CFT with discretized central charges.

To evaluate the number of e-foldings $N_e$, defining the inflaton ${\cal I}\sim b^2k_2 + b^1k_1$, one can show that (in $M_p=1$ units)
\[N_e=-\int_{{\rm in:\ Saddle\ Point}}^{{\rm fin:\ dS\ Minimum}}\frac{1}{\sqrt{\epsilon}}d{\cal I}\sim
  \frac{\sqrt{\sum_{\beta\in H_2}(n^0_\beta)^2}}{n^s\sqrt{{\cal V}}}\]
For appropriately high degree of the genus-0 holomorphic curve (usually 5 or more - See \cite{Klemm_GV}), one could choose $n^0_\beta$'s in such a way that $n^0_\beta\sim60n^s\sqrt{{\cal V}}$. This would yield the required 60 e-foldings.

\section*{Acknowledgements}

The work of AM is partially supported by a Department of Atomic Energy (Govt. of India) Young Scientist award and PS is supported by a CSIR junior research fellowship. One of us (AM) would like to thank H.Tye, D.Kabat and specially S.Sarangi for very useful clarifications.

\end{document}